# A biologically interfaced evolvable organic pattern classifier

Jennifer Gerasimov*[1], Deyu Tu[1], Vivek Hitaishi[1], Padinhare Cholakkal Harikesh[1], Chi-Yuan Yang[1], Tobias Abrahamsson[1], Meysam Rad[1], Mary J. Donahue[1], Malin Silverå Ejneby[2], Magnus Berggren[1], Robert Forchheimer*[3], Simone Fabiano*[1]

[1]Laboratory of Organic Electronics, Department of Science and Technology, Linköping University, SE-60174 Norrköping, Sweden

[2]Department of Biomedical Engineering, Linköping University, SE-58183 Linköping, Sweden

[3]Department of Electrical Engineering, Linköping University, SE-58183 Linköping, Sweden

*E-mails: jennifer.gerasimov@liu.se, robert.forchheimer@liu.se, and simone.fabiano@liu.se

**Abstract**

Future brain-computer interfaces will require local and highly individualized signal processing of fully integrated electronic circuits within the nervous system and other living tissue. New devices will need to be developed that can receive data from a sensor array, process data into meaningful information, and translate that information into a format that living systems can interpret. Here, we report the first example of interfacing a hardware-based pattern classifier with a biological nerve. The classifier implements the Widrow-Hoff learning algorithm on an array of evolvable organic electrochemical transistors (EOECTs). The EOECTs' channel conductance is modulated in situ by electropolymerizing the semiconductor material within the channel, allowing for low voltage operation, high reproducibility, and an improvement in state retention of two orders of magnitude over state-of-the-art OECT devices. The organic classifier is interfaced with a biological nerve using an organic electrochemical spiking neuron to translate the classifier's output to a simulated action potential. The latter is then used to stimulate muscle contraction selectively based on the input pattern, thus paving the way for the development of closed-loop therapeutic systems.

1. Introduction

The exponential growth in the number of and variety of data acquisition tools motivates the drive to develop neuromorphic hardware for local data storage and processing. Under the broad umbrella of neuromorphic computing, which takes inspiration from biological approaches to information processing, new devices, networks, and architectures are rapidly being developed to resolve this data handling bottleneck. The next generation of bidirectional brain-computer interfaces (BCIs) would particularly benefit from a local, reconfigurable information processing network as wireless transfer of raw data from smaller and more numerous implantable sensors becomes limiting. The ideal BCI would continuously receive input data from a panel of sensors, classify the input patterns into relevant and irrelevant events, and deliver the appropriate output only when a specific event was detected.

In a general sense, a neuromorphic computing system consists of two elements: "neurons" that integrate the inputs from a panel of other neurons and "synapses" that perform a weighting of the neuronal inputs. During the training phase, the synaptic weight, which regulates the extent to which one neuron induces the firing of another, is adjusted to optimize the performance of the network.

Today, neuromorphic hardware typically implements analog or multi-state resistive elements, primarily memristors and organic electrochemical transistors (OECTs), to represent the synaptic weight[1]. Of the many artificial synapse candidates for large-scale, local implementation, OECTs stand out due to their high number of accessible weights, low cost, printability, energy efficiency, and low operating voltages[2]. However, to retain a programmed conductance state, neuromorphic OECTs rely on either chemical[2c, 2d] or electrical ion trapping[3], making this kinetically-controlled process inherently volatile and sensitive to environmental oxygen[4].

Recently, we have introduced the evolvable organic electrochemical transistor (EOECT) as a synaptic element in neuromorphic circuitry[2f-h, 5]. Unlike a standard OECT[4a], the synaptic weight of an EOECT is controlled *in situ* by electropolymerizing additional semiconductor material within the channel, allowing for variable conductance in the equilibrated state. Thus, EOECTs have inherently longer retention times at ambient conditions than prefabricated OECTs. Further, EOECTs greatly simplify neuromorphic circuitry by eliminating the need for independent gates, insulation from the environment, and devices to limit gate current leakage. Polymerization takes place as soon as a sufficient voltage difference is applied between the gate and one, or both, of the drain and source terminals, leading to an increase in the synaptic weight represented by the device. By keeping the working voltages below the polymerization threshold, the device behaves similarly to other reported OECTs and the weight can be read non-destructively.

The EOECT overcomes many of the challenges faced by memristor- and OECT-based synapses. In contrast to a memristor, an EOECT can access many more conductance states, has a highly reproducible voltage threshold for conductance modulation, and high device-to-device reproducibility. When compared to synaptic OECTs, in addition to the aforementioned advantages in stability, networks built using EOECTs can be controlled at multiple levels because the conductance of each EOECT in an array can be updated either independently and simultaneously in a common electrolyte by inducing polymerization at the source/drain contacts or globally by inducing polymerization using the gate.

Here, we construct a biological interface that allows a medicinal leech to 'use' an electronic pattern classifier to distinguish between the letter 'T' and the letter 'J' written on a 4x4 pixel array. To do so, we develop a neuromorphic classifier that implements the Widrow-Hoff learning algorithm using the conductance state of an EOECT as a representation of synaptic weight. The output current of the classifier, pre-trained to produce a positive current in response to the 'T' input and a negative current in response to the 'J' input, is used as the input to an integrate-and-fire organic electrochemical neuron (OECN)[6], which produces voltage spikes in response to a positive current, but not when the input current is negative. The OECN interfaces with the ventral nerve cord of a medicinal leech through a flexible cuff electrode[7]. We can thus stimulate muscle contraction exclusively upon application of one of the letter inputs.

## 2. Results and Discussion

### 2.1. Classifier Implementation

Already in the 1940s when the first neuromorphic hardware circuits were suggested[8], various methods to find the values of the weights were devised[9]. In 1960, B. Widrow and T. Hoff[10] presented a protocol to generate the weight values iteratively from training samples. Their paper also described an implementation (ADALINE) which used the "memistor", a programmable three-terminal device based on electroplating copper on a graphite rod in a solution of sulfuric acid to hold the weights. This algorithm, known today as Widrow-Hoff learning or the Least Mean Square (LMS) algorithm, is favored in some applications, like adaptive noise filtering, due to its computational simplicity[11]. Since its first demonstration with ADALINE, the LMS algorithm has been implemented in hardware using memristors[12], optically-gated carbon nanotube transistors[13], and phase-change memory[14].

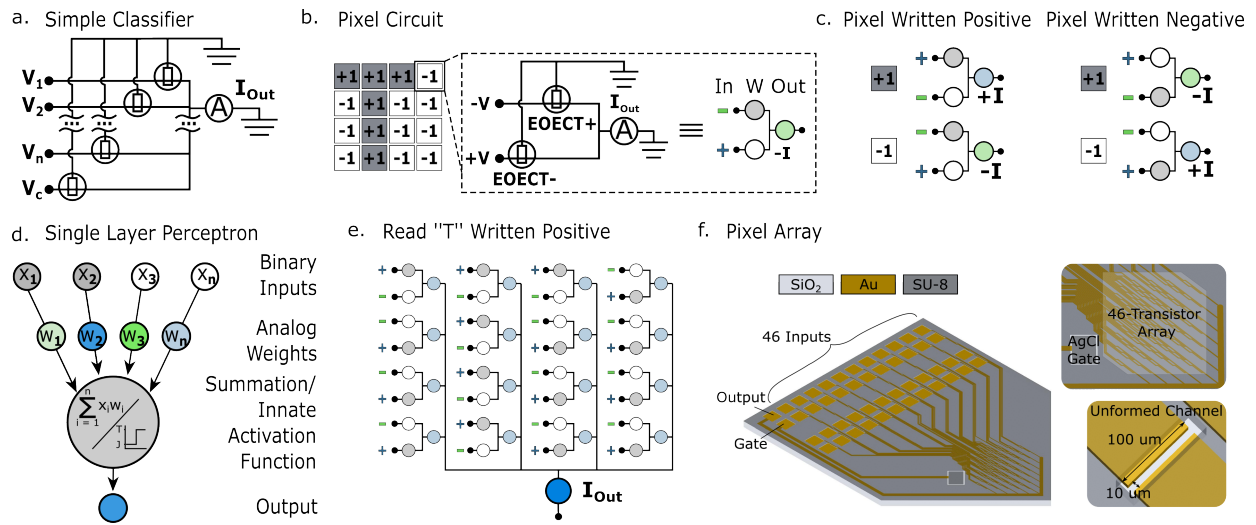

**Figure 1.** Training concept. a) Schematic of the simplest circuit producing the weighted sum of n input voltage values and implementing a thresholding operation using the control voltage $V_c$ such that the output above the threshold produces positive values and the output below the threshold voltage produces negative values. b) Schematic of a pixel where each weight is split up into a positive and a negative weight coefficient. c) All possible reading modes for a single pixel. d) Translation of the reading scheme to a single-layer perceptron e) when a pattern is written positive on a fresh array, each pixel contributes a positive current value to the output. f) Layout of the classifier used in this study.

The simplest neuromorphic EOECT network can be constructed from a panel of synapses that perform a weighting of the inputs (the "input pattern") followed by a neuron that performs a summation and thresholding, representing a classification of the input pattern into one of two classes (Figure 1a). In analogy with memory cells, we will use the terminology *write* to denote a change of a synaptic coefficient value. Similarly, by *read* we refer to measuring the output of the synapse for a given input without modification of the weights. Each transistor represents one weight, defined by the conductance of its polymerized channel. The input values $V_i$ are applied as voltages and the ammeter reads out the weighted sum as the resulting current.

Assuming that a current above a chosen threshold value represents one class while a current below the threshold signifies the second class, the network divides the input data into the two classes through a boundary defined by the weights and the threshold level. Provided that the input patterns fit into each half-plane (linearly separable patterns), the weights can be set such that the network can sort the input data into the two classes with no errors. Implementing this system with EOECTs, however, faces a major engineering obstacle that the weight can only be increased in a solution containing the monomer precursor while in order to decrease the weight, the monomer must be absent[5a]. Thus, the formation of the channels has to be based on knowledge of the final value of each weight. This seems to preclude the option to train the circuit iteratively or to adapt the weights to a change in the statistics of the input patterns.

We have resolved this problem using the "two-wire" coefficient representation. By this it is meant that the weight of each pixel is split into a positive and a negative component, each of which is controlled independently by a separate EOECT. This is accomplished by representing the input value V to the synapse in the two-wire representation as well, namely as +V and -V. The output is available as a current and is sensed by the ammeter shown in the circuit in Figure 1b. The output current from a single pixel can then be represented as:

$$I_{pixel} = (+V * G_1) + (-V * G_2) = V(G_1 - G_2)$$

where V is the magnitude of the input voltage and $G_1$ and $G_2$ are the conductances of the two EOECT components EOECT+ and EOECT-, respectively. Thus, the weight coefficient at a given pixel is represented by the difference between two positive EOECT conductance values. This representation handles bipolar coefficients and makes it possible to adjust the coefficients both up and down even though the weights can only be increased. This implementation also provides a natural boundary (0) that facilitates classification and removes the need for a control transistor to define the threshold between classes. While the two-wire representation doubles the number of devices required to produce a network with a given number of weights, it bypasses significant engineering challenges associated with microfluidic integration and allows for the future development of solid-state devices.

While it is possible to scale the input voltage to accommodate analog inputs, in this work, we use a single input voltage for the benefit of simplicity. To maintain a consistent doping level when reading positive and negative inputs, we scale the negative read input by a constant to produce an output current of the same magnitude as the positive input. In practice, this scaling factor ensures that $V(G_1 - G_2) = -V(G_2 - G_1)$. This process is discussed in more detail in the supporting information (Figure S1).

In this implementation, there are two reading modes and two writing modes (Figure 1c). At a given pixel, reading a +1 differs from reading a -1 in that the positive and negative input voltages are flipped. When reading a +1, the positive input voltage is applied to EOECT+ while the negative input voltage is applied to EOECT-. When reading a -1, the negative input voltage is applied to EOECT+ while the positive input voltage is applied to EOECT-. The writing operation also has a positive and a negative mode. When a pixel is written in the positive writing mode, the conductance of EOECT+ is increased whereas when a pixel is written in the negative writing mode, the conductance of EOECT- is increased. This has the effect that a fresh pixel, when written positive, will produce a positive output current when reading a +1 and a negative current when reading a -1. Inversely, if a fresh pixel is written negative, it will produce a negative current when reading +1 and a positive current when reading -1.

Training an array to classify a given pattern as positive, for example the 'T' pattern shown in Figure 1b, requires that all +1 pixels, defined by the pattern, are written positive and the -1 pixels are written negative (Figure 1d). In this way, each pixel will produce a positive current.

To produce a direct sum of the currents from each pixel, the outputs of the pixels are simply wired together and connected to an ammeter.

The device used to evaluate this implementation was fabricated by standard photolithography methods. Chromium-backed gold electrodes were patterned on a silicon oxide substrate and insulated with an SU-8 photoresist to define the openings for each channel. The device accommodates a 46-channel array with 46 independent inputs, although only 32 were used in this demonstration, and a single output. It should be noted that while independent gates are depicted for each EOECT in the circuit diagram in Figure 1b, a single in-plane electrode that is modified with Ag/AgCl paste is used to gate all devices within the array, simplifying the manufacturing process. The gate voltage serves to both provide stability in the doping level across all transistors and to provide the counter-reaction for monomer oxidation.

## 2.2. EOECT Characterization

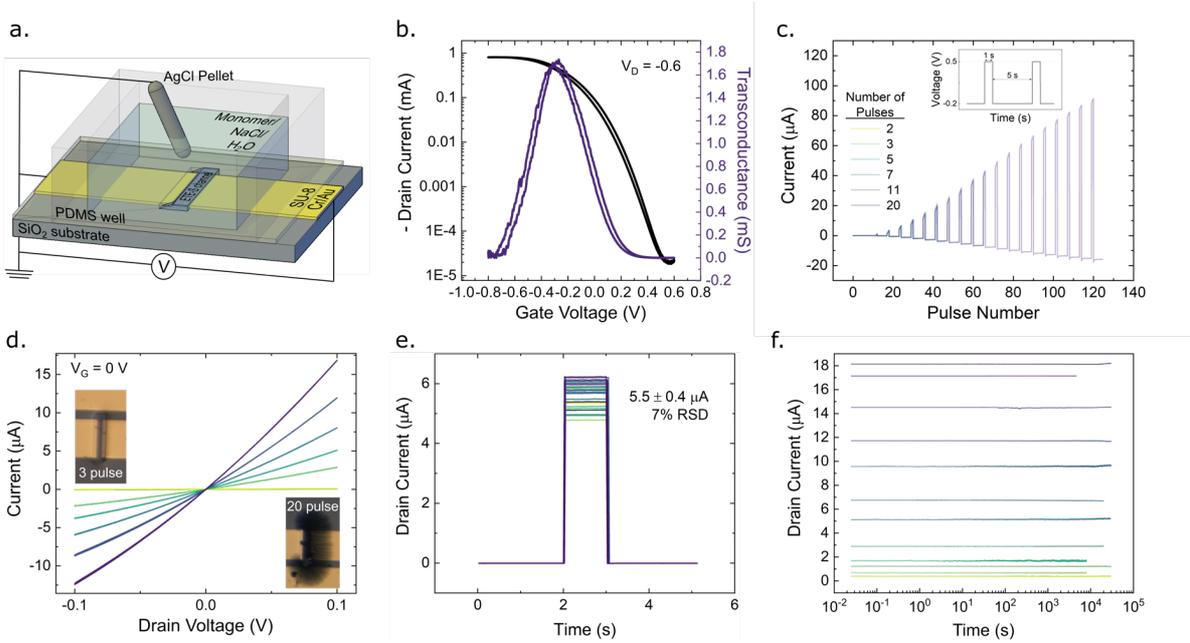

**Figure 2.** Device Characterization. a) Schematic of a single EOECT. b) Transfer characteristics of a representative EOECT. c) he time-dependent current response of six independent devices

grown by applying a sequence of one-second electropolymerization pulses (0.5 V) and characterized between pulses at a voltage of -0.2 V. A variable number of pulses, ranging from 2 to 20, was applied to each device. d) Electrical (plot) and visual (inset) characterization of the final properties of the six devices fabricated in panel c. The colors correspond to the legend in panel c. e) Current of 32 independent transistors on a single device in response to a 50 mV characterization pulse. f) Stability plots of dry, pre-fabricated devices over a wide conductance range.

A single EOECT within the array is grown by applying a voltage at the independently addressed input contact while grounding the output contact that is shared among all devices. The Ag/AgCl gate is also grounded within the reaction well to provide a counter-reaction for the monomer oxidation and to stabilize the conductance state of the semiconducting channel (Figure 2a). We use a monomer with a 2,5-bis(2,3-dihydrothieno[3,4-b][1,4]dioxin-5-yl)thiophene (ETE) backbone, which has been functionalized on the central thiophene with a sodium ethoxy-1-butanesulfonic acid sidechain (ETE-S), to form the semiconducting (PETE-S) transistor channel. In this way, the channel conductance can be monitored *in operando* during the reading and writing processes. Electrical characterization of the device reveals that electropolymerized ETE-S forms p-type devices operating in a hybrid accumulation-depletion mode (Figure 2b). More specifically, the application of a negative gate voltage results in an increase in drain current by a factor of 12 relative to 0 V while the application of a positive gate voltage can reduce the drain current by a factor of 3500. This behavior is to be expected of ETE-S, as the sulfonate sidechain serves to balance the charge of the mobile hole on the sulfonate backbone. In a previous publication, we have shown that this behavior can be useful to induce short-term, transient potentiation and depression of the synaptic weight, which will likely prove useful for the analysis of dynamic time-series data[5a].

Since this work focuses on the classification of static patterns, training through long-term changes in the channel conductance state is of greater interest. In this implementation of the LMS algorithm, we train each channel of the device using 1 s training pulses of 0.5 V, applied at the independently addressed input terminal. A sequence of training pulses separated by a 5 s refractory period of -0.2 V, was applied to each of six channels on a single device (Figure 2c). The number of pulses in the sequence ranges from 2 to 20 pulses. A plot of the current as a function of time for all six transistors shows that a continuous conducting channel consistently forms during the second pulse and increases incrementally with each successive pulse.

The IV curves around the origin, taken below the polymerization threshold, were used to quantify the conductance of each channel fabricated by pulsed ETE-S deposition (Figure 2d). The conductance ranges from 0.4 µS for the channel to which two pulses were applied to 145 µS for the channel to which 20 training pulses were applied. The channels shown in the inset microscopy images were grown by applying a voltage above the electropolymerization threshold to the electrode depicted on the left side of the image. It is apparent that an ETE-S film spreads uniformly along the substrate away from the electrode where the electropolymerization voltage is applied, except for where the polymer contacts the ground. We recently argued that this behavior is caused by the high effective concentration of the monomer near the substrate due to the adsorption of anionic ETE-S to the amine-modified surface[5c].

EOECT device-to-device reproducibility (Figure 2e) and stability in ambient conditions (Figure 2f) are especially remarkable compared to existing technologies. The currents measured during a 50 mV read pulse for devices fabricated by 5 sequential write pulses deviated by only 7 % within an array of 32 transistors. Once fabricated, the channel resistance is incredibly stable in the dry state, changing by 1% in an average of 9.5 hours, which represents an improvement of two orders of magnitude over state-of-the-art OECT devices (Figure S2)[3a].

## 2.3. Training the array

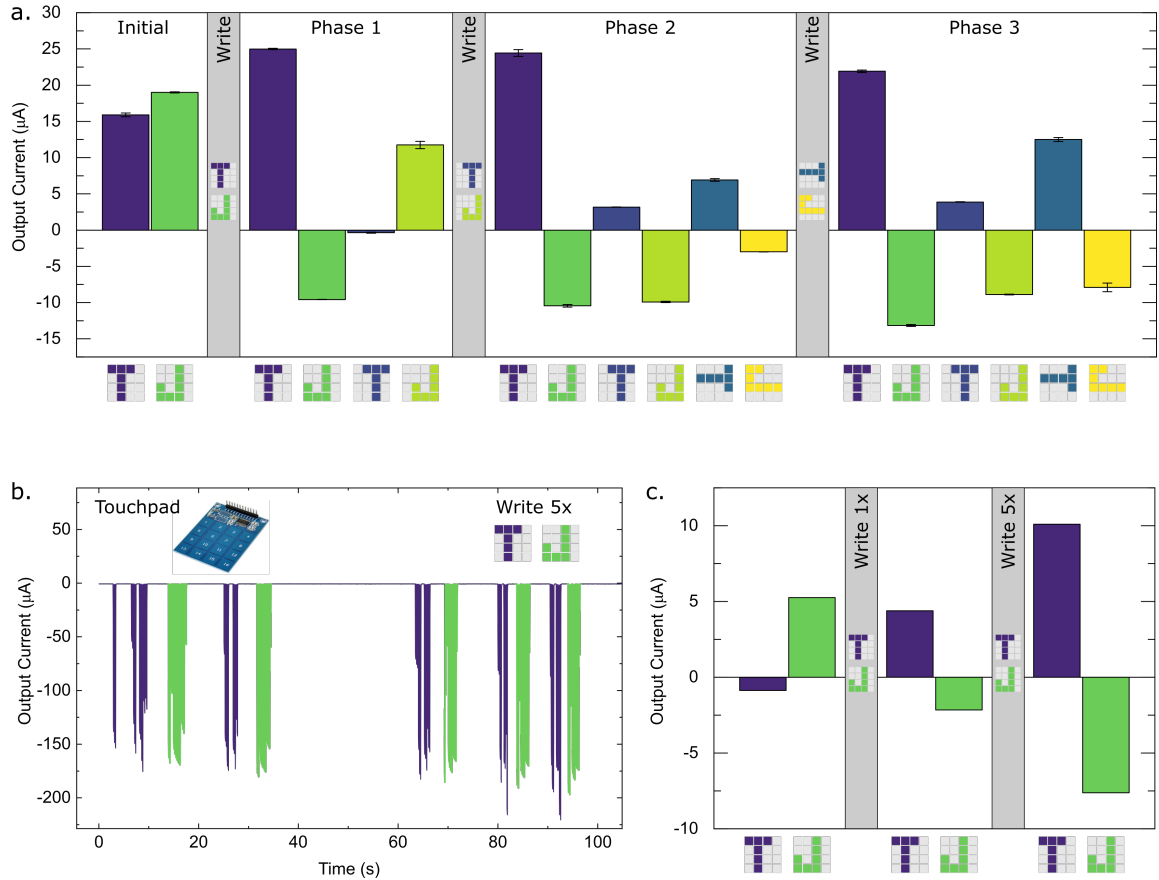

**Figure 3.** Training the array. a) The output currents of an array during training in response to a given input pattern (x-axis). The array was initialized using a random set of weights and trained before each characterization phase to produce a positive output for the 'T' input pattern and a negative output for a 'J' input pattern using the patterns in the grey blocks. The average and standard deviation represent data from four successive read operations. Training in panel a was conducted using a microcontroller. b) Training and c) characterization of the training performed using a touchpad.

Our validation of the EOECT-based approach to classify a set of input patterns closely follows the results obtained from ADALINE. In particular, we mimic a training sequence shown by Widrow[15]. The input patterns are the 4 x 4 pixel binary images shown in Figure 3. A network consisting of 16 synapses is trained to classify the letters "T" and "J". Each letter has a full height (4 pixels) and a width of 3 pixels. The letters are allowed to be in two vertical positions and one horizontal position. Altogether there are thus 6 different objects which should be correctly classified as either T or J.

The array is initialized with random weights, resulting in the correct classification of the T pattern written on the left side of the 4 x 4 array, but not of the J pattern, as reading both T and J produces a positive voltage (Figure 3a). Average values and standard deviations in this case are obtained from currents generated by four successive read operations. After a single training iteration, which consists of two one-second writing pulses, is applied to all pixels in the array, both T and J written on the left are correctly classified. In Phase 1, we also interrogate the output from T and J written to the right side of the array to find that both outputs are incorrectly classified. The second training iteration is then applied to write T and J on the right side of the array, which results in their successful classification in Phase 2. As before, we also evaluate T and J patterns that this device has not yet been trained to recognize. In this case, the patterns T and J have been rotated 90° clockwise and the device is already able to classify both of these patterns. When training is carried out for the rotated T and J in the third iteration, the current values for these patterns increase (Phase 3). It is shown here that a single training iteration is sufficient to yield the correct classification of a given pattern.

To demonstrate writing of the classifier using sensory inputs, a touchpad sensor array is coupled to the classifier via an Arduino microcontroller. Consisting of 16 capacitive keys in a 4x4 array (Figure S3), the touchpad is designed to produce a train of 100 ms spikes at a period of 200 ms for as long as each sensor is pressed down. To process the pressure signals, an Arduino microcontroller board is configured to deliver the output training signals to the synapse array. For "T" pattern, the touch keys 1, 2, 3, 6, 10, and 14 are involved and the corresponding signals are connected to the "EOECT+" devices (Figure 1b) in the synaptic pixels to increase the weight once those keys are touched. For "J" pattern, the touch keys 3, 7, 11, 15, 14, 13, and 9 transfer pressure signals to the "EOECT-" devices (Figure 1b) in the corresponding pixels to reduce the weight. The write currents generated in response to applying a series of 0.5 V spikes that are produced when letters are traced on the touchpad are represented in Figure 3b. It can be seen that the T was written as two strokes while the J was written as a single stroke on the touchpad. We again started with a randomized array that produced an incorrect classification of the T and J patterns (Figure 3c). After writing the T and J using the touchpad, the array successfully classified the two patterns. This is notable because the number of spikes used to write each pixel, which corresponds to the length of time that each button was pressed down, ranges from 1 to 7, demonstrating the robustness of array training.

**2.4. Biointerfacing**

Integrating the current or voltage output of a neuromorphic classifier circuit with biological neurons poses several challenges, including matching the voltage, current, and frequency to fall within the parameter space that is suitable for neural stimulation. As a means of bridging that gap, we couple the EOECT-based classifier that was pre-trained to distinguish between 'T' and 'J' inputs to an integrate-and-fire organic electrochemical neuron (OECN) circuit, which is then interfaced with the ventral nerve cord of a medicinal leech via a flexible cuff electrode[7] (Figure 4a). The OECN is analogous in many behaviors to biological neurons and has previously been used to induce action potentials in plants[6]. In contrast to the previously reported circuit[6], the OECN used in this work incorporates transistors with shorter channels and eschews the capacitance-voltage divider circuit, relying instead on the innate capacitance of the transistor components to serve the same function. This OECN fires at an amplitude of 0.3-0.5 V and a frequency proportional to magnitude of a positive injected current (Figure S4). If a negative current is injected, the neuron shows no response.

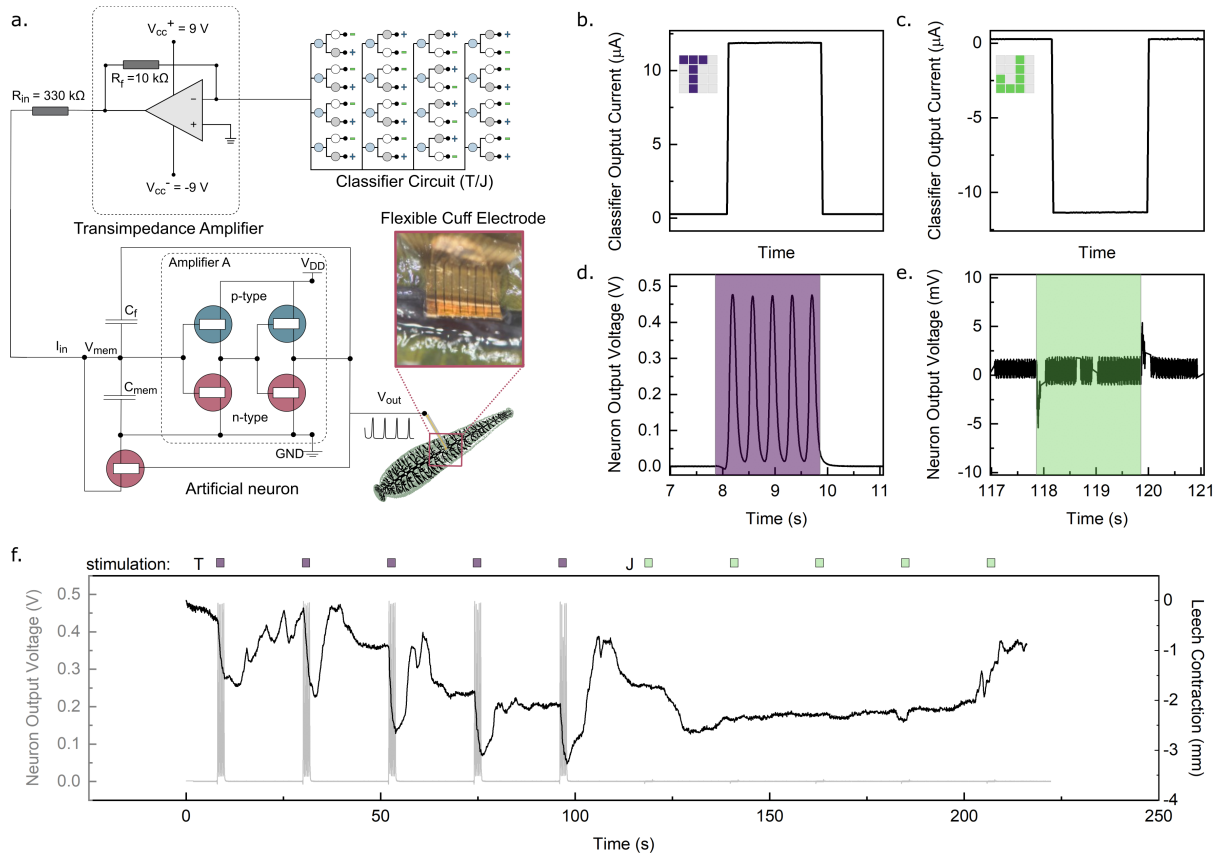

**Figure 4.** Pattern classification with a biological actuator. a) Schematic of the connection between the classifier circuit, artificial neuron, and leech. b) Output current of the pre-trained classifier to a 'T' input and c. a 'J' input. d) Output voltage from the artificial neuron in response to a 'T' input and e. a 'J' input. f) Output voltage from the artificial neuron in response to a sequence of five T and five J inputs (grey) and a simultaneously acquired estimate of leech motion (black).

When coupled to the classifier circuit through a transimpedance amplifier, OECN produces a cathodic spike train with a voltage amplitude of 0.47 V, a frequency of 2.5 Hz, and a half-width duration of 100 ms in response to a T input (Figure 4d) while producing no spiking behavior in response to the 'J' input (Figure 4e). We apply T and J inputs to the classifier for two seconds at twenty second intervals while tracking both the voltage output of the neuron and the movement of the leech, derived from video analysis (Figure 4f). Muscle contractions were evaluated by tracking positional markers on the surface of the leech. We note a significant increase in muscle contraction when the T input is applied, but not when the J input is applied. Thus, through this approach, we can enhance leeches with the ability to recognize the letter 'T' without the capacity for literacy or sight.

## 3. Conclusions

We have built a binary classifier using the Widrow-Hoff learning algorithm, wherein synaptic weight coefficients are represented by the difference in the conductance of two EOECT channels. Splitting the weight coefficient into a positive and a negative component reduces the complexity of the EOECT-based system and allows for further integration into solid-state devices. EOECTs are uniquely well-suited to hold synaptic weights in a neuromorphic hardware system due to their low operating voltages, reproducibility, and stability. We train the EOECT-based classifier to produce a positive output current in response to a 'T' input and a negative current in response to a 'J' input written on a 4x4 pixel array. We then couple the classifier to a spiking integrate-and-fire OECN that translates a positive output current into a train of voltage spikes while exhibiting no spiking in response to a negative output current. The output of the OECN is then delivered directly to the ventral nerve cord of a medicinal leech via a flexible cuff electrode. Thus, the leech exhibits muscle contraction when a 'T' input is applied and no contractions when a 'J' input is applied. This work represents a breakthrough in the effort to integrate local decision-making hardware with biological systems.

## 4. Experimental Section

### 4.1. Instrumentation

*Voltage pulse generator:* A programmable 48-channel voltage pulse generator was constructed using three 16-channel evaluation boards coupled to a myRIO device (Figure S5). A custom LabVIEW program was used to define the parameters of the voltage pulse sequence generated by the device.

*Touchpad setup*: a TTP229 (TonTouch) 16-key capacitive touch keypad from TonTek is interfaced with an Arduino Mega 2560 microcontroller (ATmega 2560) board, shown in Figure S3. The Arduino Mega 2560 is programmed to take inputs from the TTP229 touch pad and deliver output pulse trains (5V) to the synapse array through resistive voltage dividers (82k$\Omega$/10k$\Omega$).

*Transimpedance amplifier*: a µA741 operational amplifier from Texas Instruments is configured as a transimpedance amplifier to bridge the classifier circuit (reading at 50 mV) and the OECN artificial neuron (firing at 0.3-0.5 V), shown in Figure S6.

### 4.2. Fabrication

*Classifier Array Fabrication:* Four-inch silicon wafers with 20 kÅ thermally grown oxide (Silicon Valley Microelectronics, Inc.), stored in a class 1000 clean room environment, were used without further treatment aside from removing dust with a stream of nitrogen. A 2 nm chromium sticking layer and an 18 nm gold film were deposited on the wafers by thermal evaporation (Moorfield Nanotechnology evaporator). Positive photoresist Microposit S1805 (MicroChem) was patterned on the metal layers by photolithography according to specifications provided by MicroChem using a Karl Suss MA/BM 6 mask aligner, after which the metal layers were etched using an $I_2$/KI solution for Au, and a ceric ammonium cerium (IV) nitrate-based etchant for chromium (Merck Life Science AB). The photoresist was then removed by rinsing in MicroChem Remover 1112A, after which the substrate was rinsed with water and acetone, dried under vacuum at 130 °C for 5 minutes, and exposed to oxygen plasma (200W, 5 minutes). Active areas of the source and drain electrodes were defined by a 5-micron layer of SU-8. The SU-8 3010 (MicroChem) photoresist was modified by diluting 5 parts SU-8 with 1 part cyclopentanone. After patterning, wafers were stored in ambient conditions for future use. An image of the finished wafer is shown in Figure S7.

Within a month before use, devices were dried by heating for 10 minutes on a hot plate pre-heated to 130°C, cleaned by oxygen plasma for 5 minutes at a power of 50 W, and modified with aminopropyl triethoxysilane (APTES). Gas-phase APTES modification was carried out by placing devices in a 130-mm stainless steel petri dish, dropping 70 µL of APTES on the bottom of the dish around the substrates, closing the lid, and incubating at 80°C for 2 hours in a fume hood. After 2 hours, devices were sonicated in 2% Hellmanex at 60°C for 5 minutes, rinsed three times with DI water and then sonicated in DI water for 5 minutes, followed by a sonication in acetone for 5 minutes and a final sonication in isopropanol for 5 minutes.

ETE-S monomer was synthesized by following previously reported protocols.[5b] PETE-S channels were fabricated by applying a voltage of 0.5 V to the independently addressed inputs of the array in the presence of an ETE-S monomer solution containing 10 mM NaCl.

*Neuron Fabrication:* The organic electrochemical neuron is fabricated following a procedure reported previously. [6, 16] Standard microscope glass slides were cleaned via successive sonication in acetone, deionized water, and isopropyl alcohol, and dried with nitrogen. Source/drain electrodes (5 nm Cr and 50 nm Au) were thermally deposited and photolithographically patterned by wet etching. A first layer of parylene C (4 µm), deposited together with a small amount of 3-(trimethoxysilyl)propyl methacrylate (A-174 Silane) to enhance adhesion, acted as an insulator to prevent disturbing capacitive effects at the metal liquid interface. Subsequently, a dilution of industrial surfactant (2% Decan-90) was spin-coated as an antiadhesive layer and a second layer of parylene C (4 µm) was deposited as a sacrificial layer. To protect the parylene C layers from a subsequent plasma reactive ion etching step (150 W, O2 = 500 sccm, CF4 = 1000 sccm, 510 s), a thick positive photoresist (10 µm, AZ10XT520CP) was spin-coated on the parylene C layer. A second photolithographic patterning step was carried out to define the contact pads and the OECT channel, and the AZ developer was applied to the photoresist. The subsequent plasma reactive ion etching step was used to indiscriminately remove the layer of organic materials, including both photoresist and parylene C, so that the contact pads and the channel area were exposed to the air while other areas were still covered with two layers of parylene C. The channel between were patterned to obtain W/L = 400 µm/40 µm for p-type OECT and 1600 µm/40 µm for n-type OECT. P($g_4$2T-T) was dissolved in 1,2-dichlorobenzene (ODCB) at 100 °C for 12 h to obtain P($g_4$2T-T)-ODCB solution with a concentration of 3 mg mL−1. The solution was spin-cast (2000 rpm, 90 s, acceleration 2000 rpm s−1) to obtain P($g_4$2T-T) thin films. The sacrificial layer was peeled off and the P($g_4$2T-T) film on it was removed, leaving separated pieces of film staying in the wells, consisting of the semiconductor connecting the OECT source/drain electrodes. BBL film was fabricated using a similar procedure through spin-casting BBL-MSA solution and parylene C patterning. The Ag/AgCl paste (Sigma-Aldrich) was drop-casted to form the gate electrodes. For all OECTs, 0.1 m NaCl aqueous solution was used as the electrolyte. The organic electrochemical neuron is assembled by connecting two P($g_4$2T-T)-based OECTs and three BBL-based OECTs using silver paint[6].

*Flexible Cuff Fabrication:* A flexible electrode array was fabricated to interface with the ventral nerve cord to provide stimulation recording capability. This array consists of 8 electrodes with dimensions of 450 µm x 200 µm in a 2 x 4 arrangement. Flexible parylene C substrate and insulation layers are used allowing the array to be used in a 'cuff-style,' wrapped conformally around the nerve. the device was created through microfabrication techniques previously described[7] to create a cuff-like layout that can conform closely around the nerve. Briefly, a 2 µm thick, flexible parylene C layer is deposited by CVD (Diener electronic GmbH) on glass microscope slides cleaned by ultrasonication first in 2% Hellmanex in deionized water (DIW), followed by acetone, and then isopropanol. Using a negative photoresist (AZ nLof 2070), a MA6 Suss mask aligner with i-line filter, and developer (AZ 326 MIF), 80 nm thick gold interconnects with a 5 nm titanium adhesion layer are photolithographically patterned. The metal interconnects allow electrical connection between the electrodes and the back.end contact pads, and provide a means for wiring upon completion. Next, a 2 min O2 plasma process is carried out (50 W) followed by deposition of a 1.5 µm thick insulating PaC layer over the metal lines (with an adhesion promoter in the deposition chamber, A-174). To define the outline shape of the probes, reactive ion etching (RIE) is carried out (O2/CF4 gases, 150 W) after a photoresist etch mask (AZ 10XT) is patterned. The photoresist etch mask is removed with an acetone wash and isopropanol rinse. A dilute (2.5 vol % in DIW) soap anti-adhesion layer is spin coated (1000 rpm) onto the sample surface. A 2 µm thick sacrificial PaC layer is then deposited. Afterward the same RIE etch process is again used with the AZ 10XT photoresist etch mask to define the electrode surface and to provide back-end contact possibility. A PEDOT:PSS-based dispersion (CleviosTM PH 1000 from Heraeus Holding GmbH), 5 wt % ethylene glycol, 0.1 wt % dodecyl benzene sulfonic acid, and 1 wt % of (3-glycidyloxypropyl)-trimethoxysilane (GOPS)) is utilized at the electrode sites to reduce electrochemical impedance. This formulation was spin coated onto the substrates, baked at 100 C for 60 seconds, and the sacrificial PaC layer is removed through a peel-off technique. The samples are annealed at 140 C for 45 minutes to crosslink the conducting polymer film. DIW is used to assist in the removal of the completed probes from the glass substrates.

### 4.3. Device Characterization

*EOECT Characterization:* Output and transfer characteristics, reproducibility, and stability studies were carried out using a Keithley 2614B Source Measure Unit.

*Neuron Characterization:* The organic electrochemical neuron was characterized using a Infiniium 54830 Series Oscilloscopes (Agilent Technologies) at a range of input currents.

*Cuff characterization:* Prior to interfacing the OECN with the nerve, stimulation of the leech nerve was performed using the Intan RHS 128-channel stimulation/recording controller and software (Intan Technologies, CA, USA) for delivered charge known to elicit contractions in the leech (24 µA for 266 µs pulse duration applied to all 8 electrodes simultaneously, see Fig SX). In all cases, a zero insertion force clips (ZIF-Clip) on custom printed circuit board (PCB) adapters were used to connect to the electrode arrays. When using the Intan controller for stimulation, this ZIF clip and PCB provided connection to a 16-channel headstage. Similarly, when interfacing the OECN to the leech nerve, the custom adapters provided wiring capability.

### 4.4. Array training

Both writing and reading are performed using a MyRIO microcontroller coupled to three 16-channel evaluation boards to produce 48 functional channels, of which only 32 were used in this experiment. A LabVIEW program is used to control the MyRIO to apply the writing and reading voltage pulses. The output current from each voltage pulse was quantified using a Keithley 2614B System Source Measure Unit to produce the values depicted in Figure 3 and the standard deviation for each value is calculated from three successive reading pulses. Each iteration of writing is composed of one-second pulses of 0.5 s.

**4.5. Integration of the classifier and OECN**

The OECN can spike at frequencies in the range of 1 to 2.5 Hz for input currents in the range of 0.5-5 µA. To bring the classifier output voltage to the operating range of the neuron, a transimpedance amplifier circuit is used (Figure S6) as a bridge. Hence the OECN can spike at 2.5Hz for the high positive current output from the classifier while remaining quiescent for negative outputs.

**4.6. Biointerfacing**

Adult leeches (Hirudo medicinalis, large, in quarantine for 32 weeks) were obtained from Biebertaler leech breeding farm (Biebertaler Blutegelzucht GmbH, Biebertal, Germany). Quality assurance requirements from the breeder were followed to store the leeches. This included the use of plastic bottles two-thirds filled with tap water (Norrköping, Sweden). The water was changed every second to third day and the bottles with leeches were stored in dark at room temperature (20-23 °C). Prior to experimentation, the leeches were anesthetized by submerging in crushed ice for 15-20 min. Subsequently, the head ganglion and sucker were immediately removed. The resulting semi-intact preparation was stretched out and pinned down with the ventral side up. A ~1.5 cm long incision was made along the midline to expose the connecting nerve. On the side of the incision, the muscles and bilateral nerves were kept intact. Before wrapping the flexible electrode array around the nerve, the black stocking surrounding the nerve was removed between two ganglions. The semi-intact leech preparation was kept wet from underneath with ice-cold solution throughout the experiment (115 mM NaCl, 4 mM KCl, 1.8 mM $CaCl_2$, 10 mM HEPES with the pH was set to 7.4 using NaOH). Movements of the leech resulting from stimulation of the connective nerve and induced muscle contractions close to the incision site were filmed with a Canon Legria Hf R86 video camera. Bright- colored analysis markers were attached to the surface of the leech on either side of the to assist in movement tracking. Movements from the video were tracked and quantified using the Tracker video analysis and modeling tool. The change in the distance between the two markers (relative to the initial distance) is used to estimate muscle contraction.

**Acknowledgments**

The authors thank Dr. Philipp Kühne and Robert Huisman for thought-provoking discussions and experimental support. This project was financially supported by the Knut and Alice Wallenberg Foundation, the Swedish Research Council (2018-06197, 2020-03243), VINNOVA (2020-05223), the Swedish Foundation for Strategic Research (RMX18-0083), the European Research Council (834677 "e-NeuroPharma" ERC-2018-ADG), and the Swedish Government Strategic Research Area in Materials Science on Functional Materials at Linköping University (Faculty Grant SFO-Mat-LiU 2009-00971).

**Data availability**

The data that support the findings of this study are available from the corresponding authors upon reasonable request.

**Competing interests**

The authors declare no competing interests.

Supporting Information

# A biologically interfaced evolvable organic pattern classifier

Jennifer Gerasimov*[1], Deyu Tu[1], Vivek Hitaishi[1], Padinhare Cholakkal Harikesh[1], Chi-Yuan Yang[1], Tobias Abrahamsson[1], Meysam Rad[1], Mary J. Donahue[1], Malin Silverå Ejneby[2], Magnus Berggren[1], Robert Forchheimer*[3], Simone Fabiano*[1]

[1]Laboratory of Organic Electronics, Department of Science and Technology, Linköping University, SE-601 74 Norrköping, Sweden

[2]Department of Biomedical Engineering, Linköping University, SE-581 83 Linköping, Sweden

[3]Department of Electrical Engineering, Linköping University, SE-581 83 Linköping, Sweden

*E-mails: jennifer.gerasimov@liu.se, robert.forchheimer@liu.se, and simone.fabiano@liu.se

**The Least Mean Squares Algorithm**

The purpose of the LMS algorithm is to minimize the error between the class labels (often chosen as +1 and -1) and the weighted sums ("regression values") generated by the network. The expectation is that this procedure should also minimize the classification error rate when the threshold circuit is applied. This is not always the case as the two criteria are different. However, the LMS optimization will converge also for patterns which are not linearly separable. This is in contrast to the Perceptron learning algorithm which is based on minimizing the error rate [3].

The error that is considered in the LMS algorithm is the sum of the squared differences between the regression values $r_j = \sum_i w_i V_{ij}$ and the class labels:

$$e^2 = \sum_j (r_j - l_j)^2 \qquad (1)$$

where wi is the i:th weight, Vij is the i:th component of a labelled input pattern Vj (being part of the "training set") and lj is its label. As is seen the error is a quadratic function. Thus, the weights that minimize the error can be solved for analytically. This is a tedious procedure however that requires solving a linear equation system with as many equations as there are weights. It also requires that all the training samples are available simultaneously.

In contrast, the LMS algorithm iterates over the training patterns to approach the correct solution. Not only is the computational burden distributed over time, but the procedure allows that new patterns and/or changed labelling can be introduced to facilitate adaptivity. The basic idea is the following.

For each training pattern the weight vector W is adjusted to bring the regression value rj closer to the label associated with that pattern. This can be expressed as:

$$W[j+1] = W[j] + step * (l_j - r_j) * V_j \qquad (2)$$

where step controls the convergence rate. As an example, consider a case where the label is +1 for a certain input pattern while the regression value is less than 1. Then, all weights which are multiplied by a positive input should be increased while those which are multiplied by a negative input should be decreased. Altogether, this will make the new regression value come closer to +1. A similar procedure is done for an input pattern which is labelled -1. But here, the adjustment of the weights will be reversed. The amount which is added to each weight should be small enough not to cause overshoots but rather a smooth convergence. Following Widrow's ADALINE implementation[1] we will only consider binary input patterns. Each pattern component is assumed to have the value 1 or -1.

**Read voltage scaling**

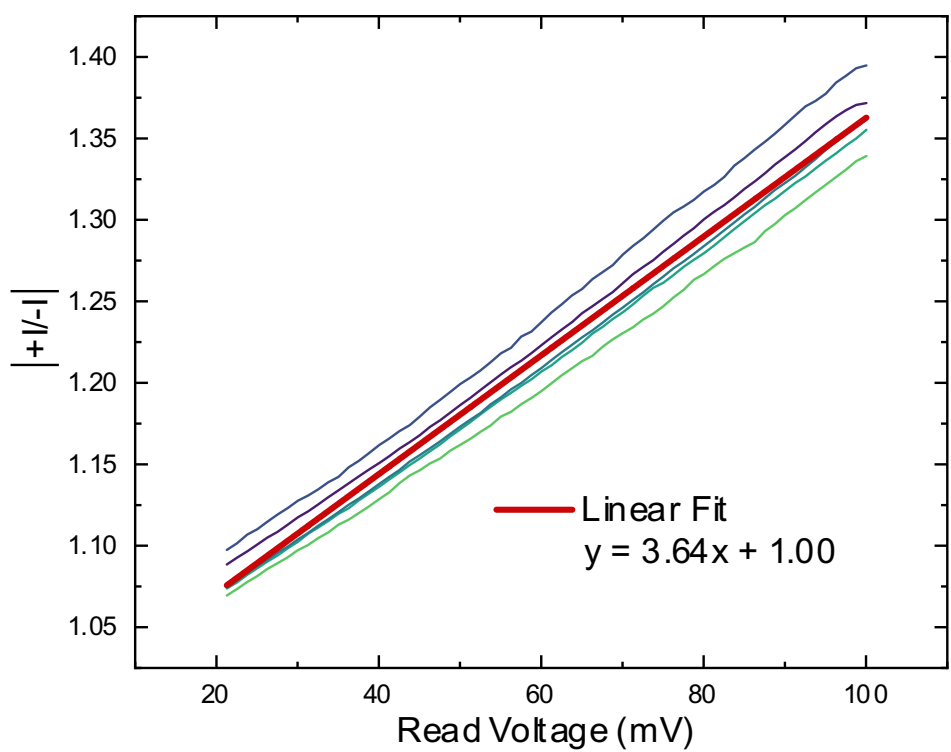

**Figure S1.** Current ratio between positive and negative voltages for a set of devices of varying conductance. Data taken from Figure 2e.

A non-destructive read can be obtained by keeping potential differences between the source/drain terminals and the gate well below 0.2 V. However, in read mode, it is very important that the positive and negative components of a given pixel are modulated so that only the difference in the effective channel volume rather than any difference in the redox state of the ETE-S channel is measured by the amp-meter. This is obtained by introducing an offset coefficient by which we multiply the negative input voltage to compensate for the reduction in the doping level. The offset coefficient is calculated from the IV data in Figure 2e by normalizing the positive drain current by the negative drain current for all drain voltages and extracting a linear fit that best accommodates all conductance states. From the fit, we can find $-V = +V \left| \frac{+I}{-I} \right|$ to calculate the appropriate negative read voltage $-V$.

**State Retention**

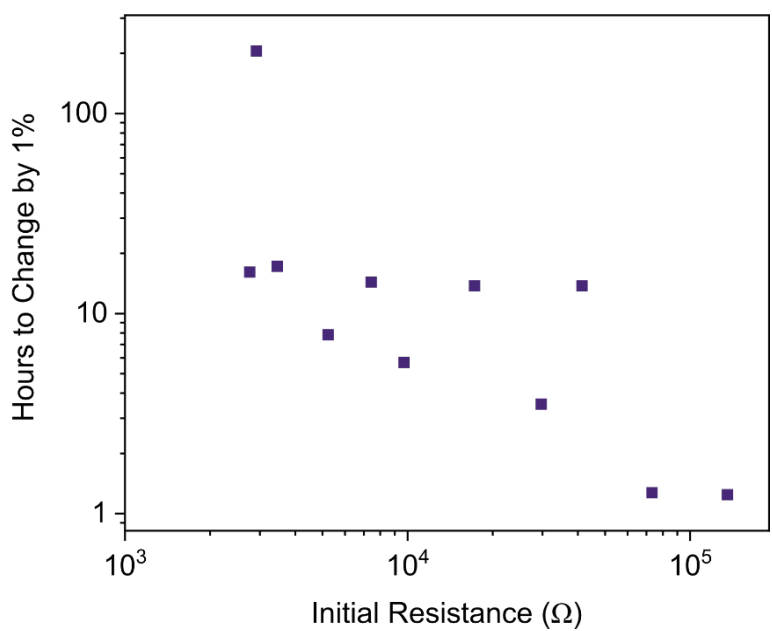

**Figure S2.** Time required for the current to change by 1%, as a function of the channel resistance. The highest point (206 hours) was taken as an outlier to obtain an average time of 9.5 ± 6.3 hours.

**Touchpad setup**

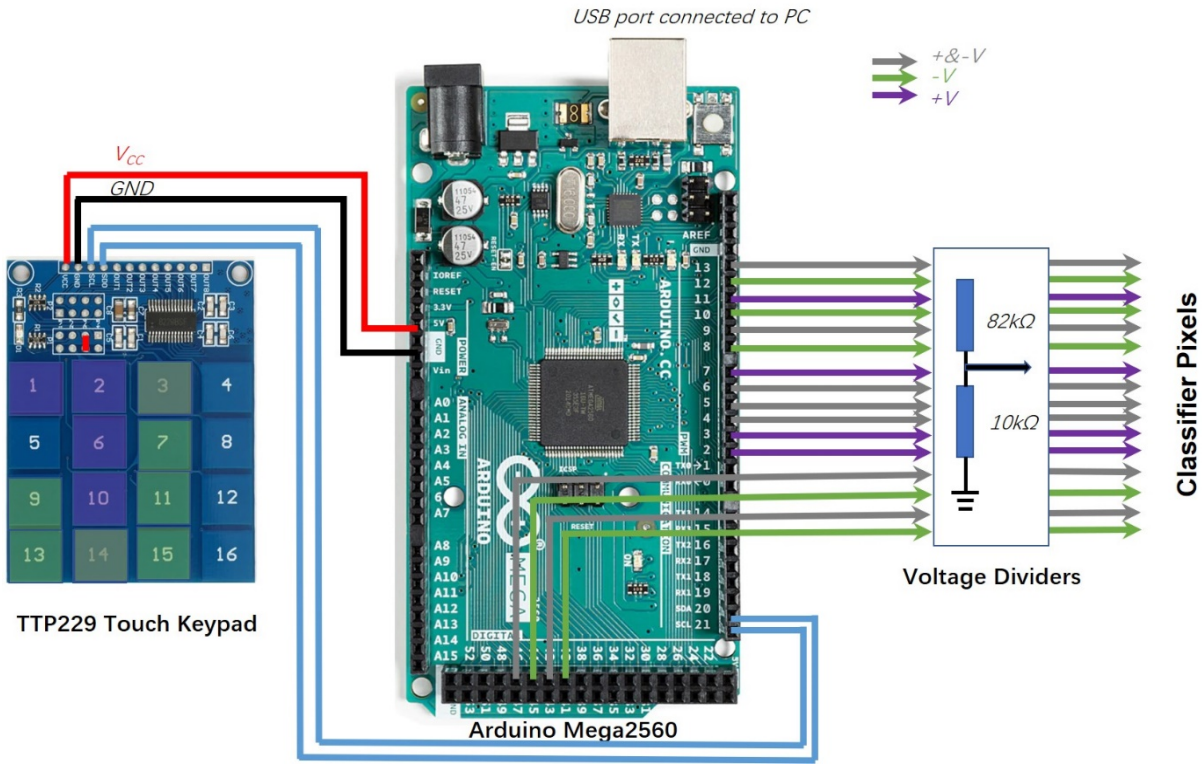

**Figure S3.** The touchpad setup for the "T/J" classifier training. The Arduino Mega 2560 microcontroller board is used to sample the touching signals from the keypad and deliver training pulses to the classifier through the voltage dividers.

**OECN characterization**

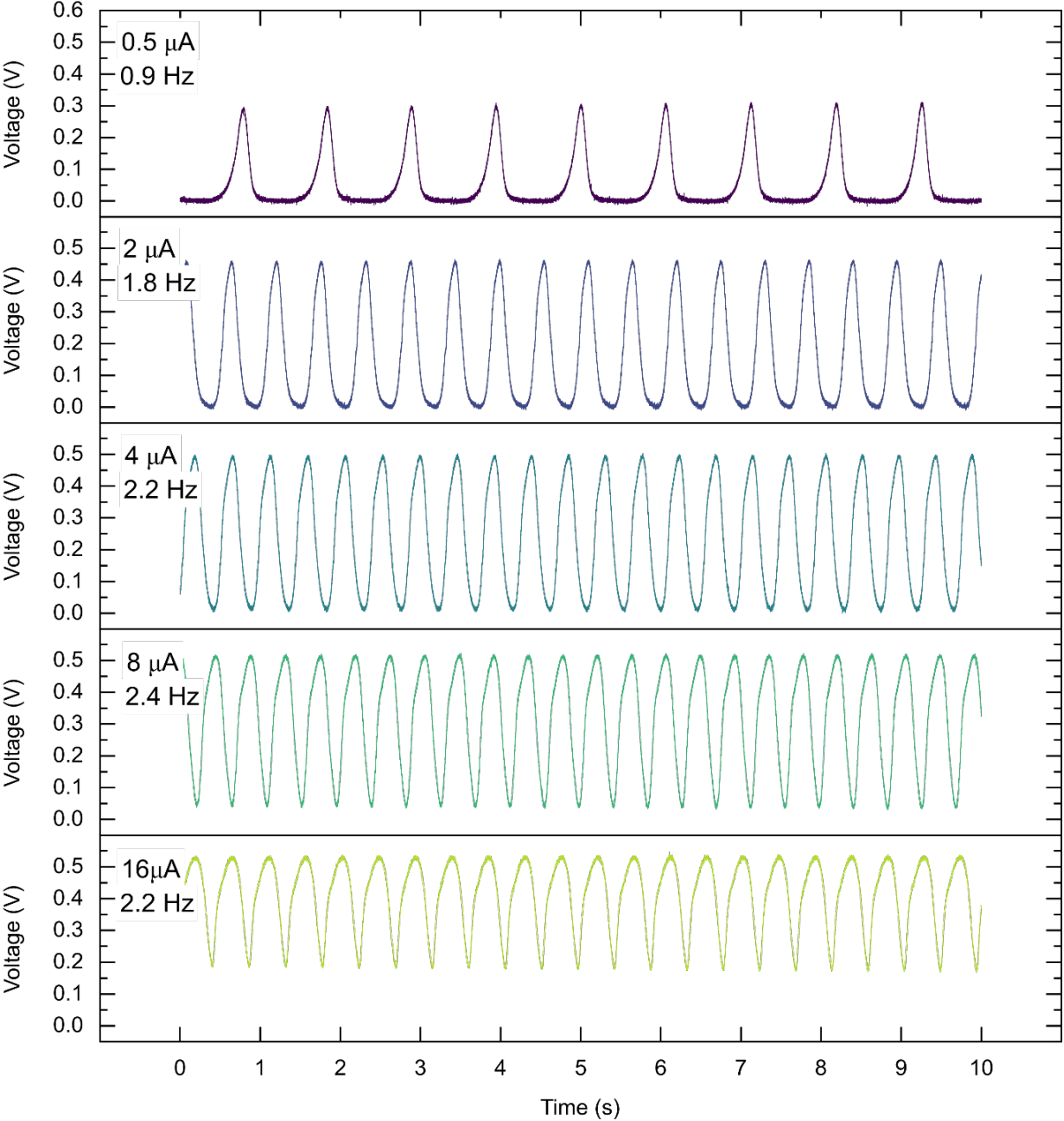

**Figure S4.** Neuron spiking behavior at a range of input currents.

**Classifier Measurement Setup**

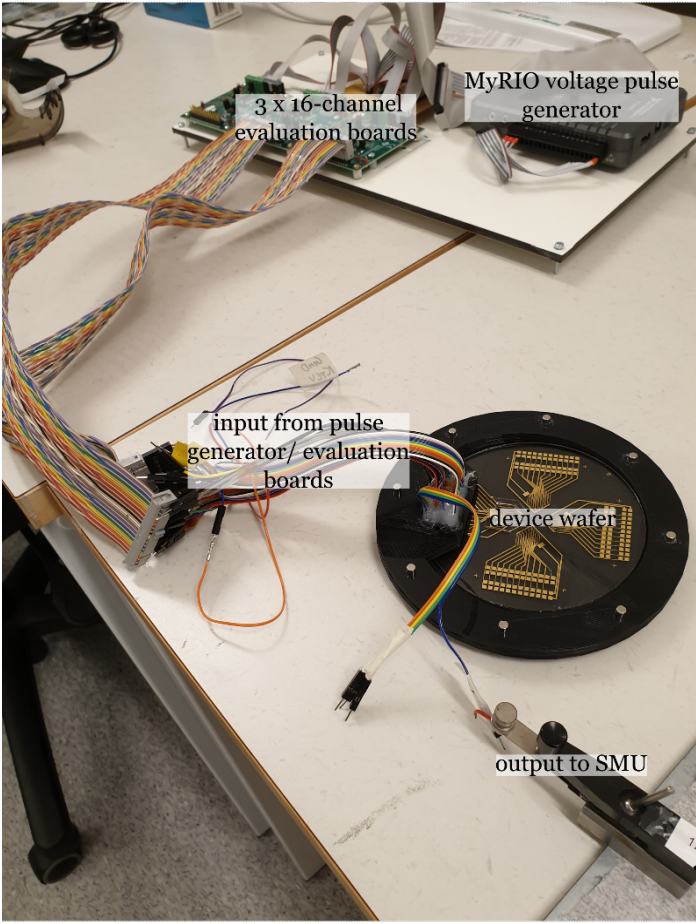
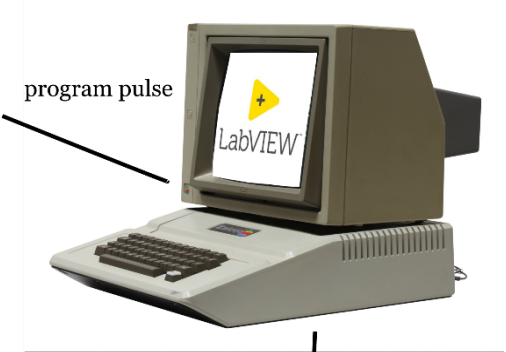
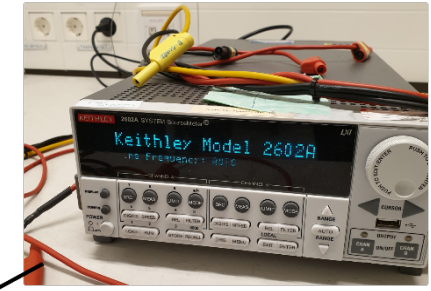

**Figure S5.** Classifier measurement setup.

**Transimpedance Amplifier**

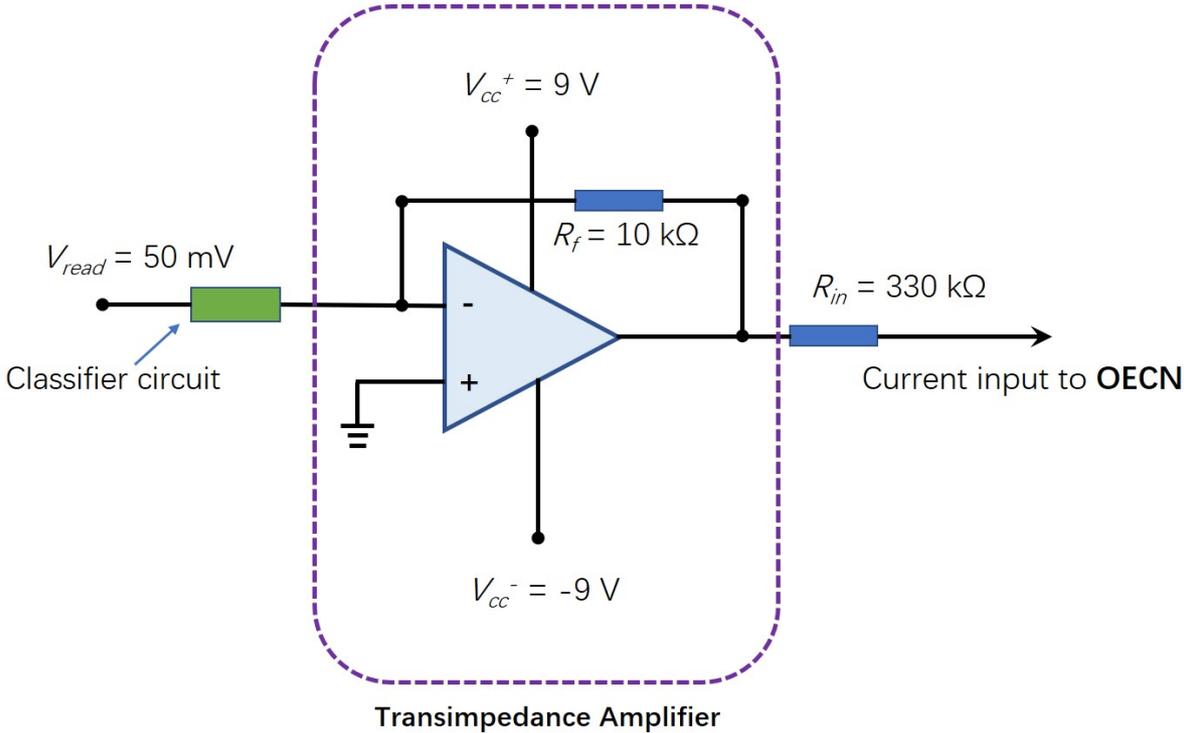

**Figure S6.** A transimpedance amplifier is used to interface the classifier and the OECN artificial neuron. Depending on the reading of the classifier, the amplifier delivers either a positive or negative current to the OECN.

**Classifier Wafer Layout**

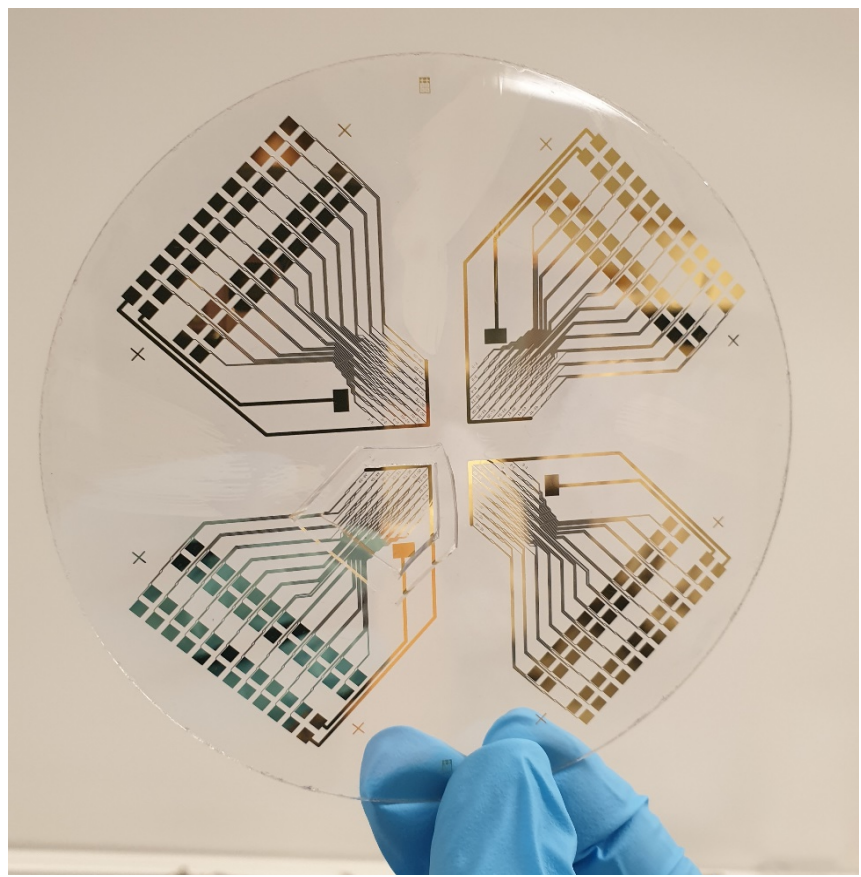

**Figure S7.** Glass wafer patterned with four EOECT classifiers.

**Current drift on positive gate voltages**

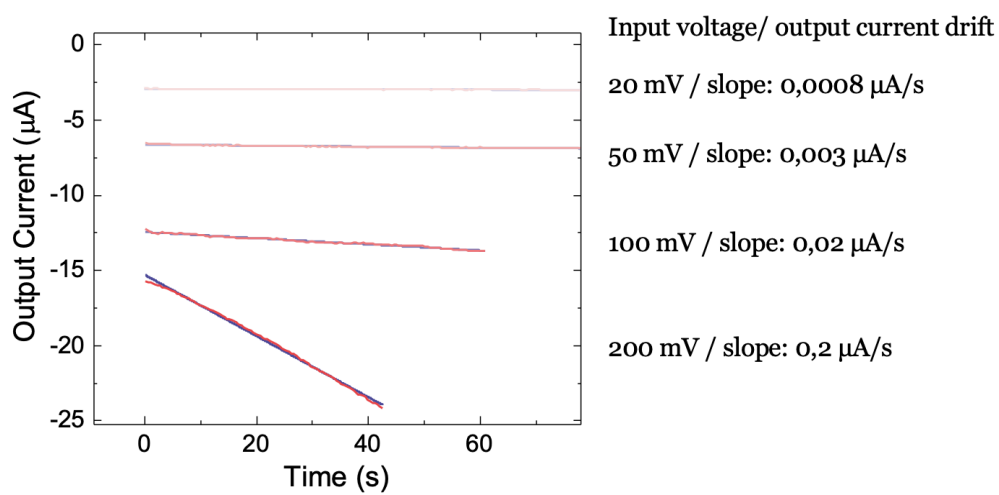

**Figure S8.** Evaluation of the drift upon the application of a positive read voltage in a solution of 1 mg/mL ETE-S with an applied gate voltage of 0 V.

**Pixel Cycling**

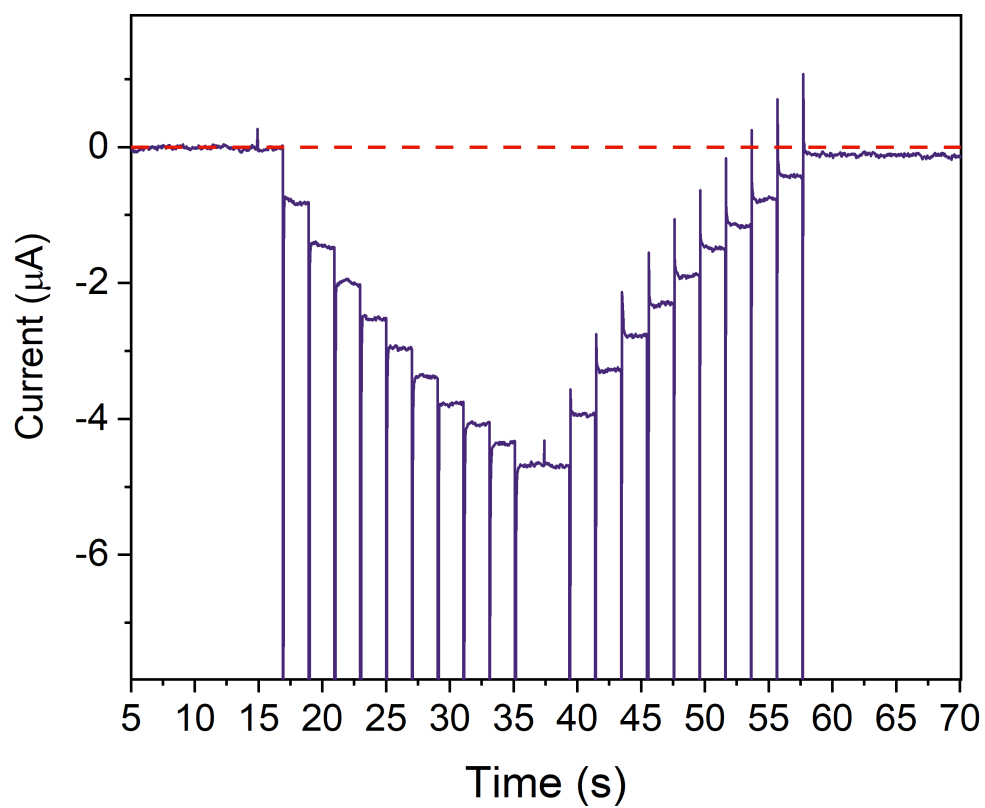

**Figure S9.** Pixel cycling. A single pixel was evaluated as the weight was bidirectionally modulated by first changing the conductance of the negative coefficient transistor with a series of 10 growth pulses and then changing the conductance of the positive coefficient transistor with a series of 10 growth pulses. Each growth pulse of 0.5V was applied for 100 ms. A linear baseline was subtracted from the data to account for baseline drift due to evaporation.

**Output Curves**

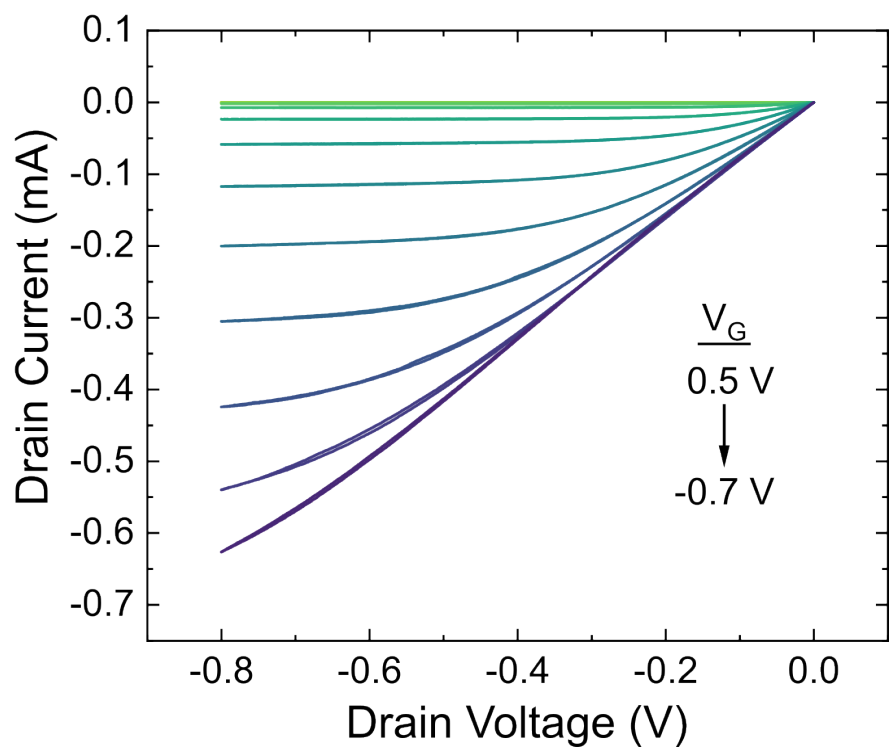

**Figure S10.** Output characterization of the device in Figure 2b, acquired in a solution of 100 mM NaCl.